# Thermodynamic limit of solar to fuel conversion for generalized photovoltaic-electrochemical system

M. Tahir Patel[1], M. Ryyan Khan[1], and Muhammad A. Alam[1, a)]

[1]Electrical and Computer Engineering Department, Purdue University, West Lafayette, IN-47907, USA

Variability of the energy output throughout the day/night poses a major hurdle to the widespread adoption of photovoltaic systems. An integrated photovoltaic (PV) and electrochemical (EC)-storage system offers a solution, but the thermodynamic efficiency ($\eta_{sys}$) of the integrated system and the optimum configuration needed to realize the limit are known only for a few simple cases, derived though complex numerical simulation. In this paper, we show that a simple, conceptually-transparent, analytical formula can precisely describe the $\eta_{sys}$ of a 'generalized' PV-EC integrated system. An M-cell module of N-junction bifacial tandem cells is illuminated under $S$-suns mounted over ground of albedo $R$. There are $K$-EC cells in series, each defined by their reaction potential, exchange current, and Tafel slope. We derive the optimum thermodynamic limit of $\eta_{sys}(N, M, K, R, S)$ for all possible combinations of a PV-EC design. For a setup with optimal-$(M, K)$ and large $N$, under 1-sun illumination and albedo = 0, the ultimate limit for $\eta_{sys} \sim 52\%$. The analysis will unify the configuration-specific results published by diverse groups worldwide and define the opportunities for further progress towards the corresponding thermodynamic limit.

## 1. Introduction and background.

Solar energy is one of the most prominent and sought after renewable sources of energy. There has been extensive investment, research and development in this field with the objective of maximizing efficiency and output power from solar cells (at device-level)[1] and solar farms (at system-level). Regardless of the aforementioned efforts, there exists a fundamental issue with solar energy viz. the source of energy (intensity of sunlight) varies within a day, with different seasons in a year as well as with latitude. Storing solar energy in various other forms of energies provides a solution to this challenge. A variety of storage solutions have been proposed, including batteries, organic and/or inorganic reactions, artificial photosynthesis[2], etc.

In this paper, we report the thermodynamic performance limit of conversion of solar energy to chemical energy (as fuels for e.g. H$_2$ and O$_2$). A generalized configuration for such PV-to-EC conversion (relevant for research efforts worldwide) is shown in Fig.1. This system is characterized by five PV and four EC variables. The PV variables are the concentration of light from sun ($S$), the fraction of incident light reflected from the ground (albedo, $R$), number of series-connected cells in the PV module ($M$), number of subcells in a multi-junction (Tandem) solar cell ($N$) and the set of bandgaps of the solar cell ($[E_g]$)). And, the EC parameters are the number of series-connected electrochemical cells ($K$), the thermodynamic potential of the reaction ($\mu_{th}$), effective exchange current density ($J_0$) and effective Tafel slope ($\beta$). An extensive literature survey[2–13] shows that the systems considered in previous works deal with only limited subset of these parameters, and the thermodynamic limits ($\eta_{sys}$) of these specific cases are computed by numerically complex simulations with results that may not always be physically transparent. In contrast, here we develop an analytical model that describes the optimum combination of parameters required to maximize efficiency of the generalized PV-EC integrated system in an intuitively transparent form, unifies the results from different configurations explored by various groups, and suggests opportunities for significant improvement by using newly developed bifacial tandem cells.

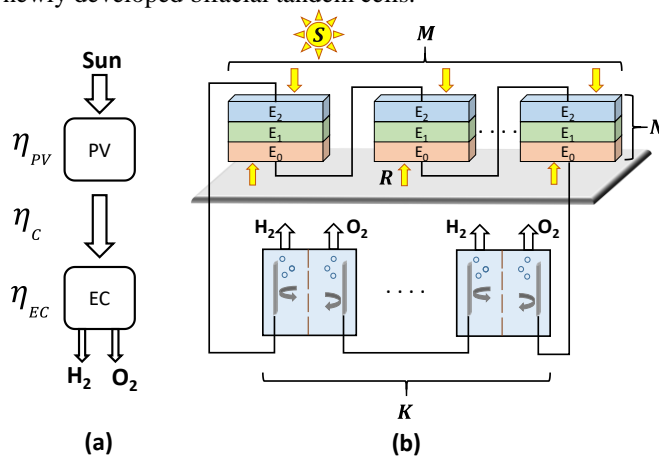

Fig. 1: (a) Energy flow diagram of a general PV-EC system (b) Schematic of the integrated system photovoltaic cell (module of tandem solar cells) - electrochemical cell

PV-EC systems may be configured in different ways[3,5,12]. Here, we focus on approach called a PV-Electrolyzer design which comprises of two *independent* pieces, namely photovoltaic (PV) cell and electrochemical (EC) cell, see Fig. 1(a). The photovoltaic part converts solar energy into electrical energy, which is supplied to the electrochemical cell that further converts this electrical energy into chemical energy in the form of fuels. This design, with physically independent pieces, is more stable and reliable with respect to material degradation as opposed to the designs where the PV parts are in physical contact with the electrolyte in the electrochemical cell[5,12]. The fuels generated from cathode and anode are then stored in containers or passed onto another electrochemical system to produce other chemicals. We assume that PV is directly connected to EC, so as to avoid losses associated with power electronics based couplers. Thus, the overall efficiency ($\eta_{sys}$) is a product of PV efficiency ($\eta_{pv}$), EC efficiency ($\eta_{ec}$) and coupling efficiency ($\eta_c$), see Fig. 1(a). We will now calculate the individual efficiencies so as to maximize $\eta_{sys}$.

## 2. I-V Characteristics and efficiencies of PV & EC Systems.

In order to find the efficiencies of the PV and EC systems, we first describe their $I-V$ characteristics in sections A and B. Then, we find an optimum point of operation so as to obtain the optimal system parameters to achieve the highest system efficiency.

### A. I-V Characteristics of an EC system.

The generalized electrochemical system comprises of '$K$' electrochemical cells connected in series. A single cell consists of two electrodes, a solution (electrolyte) and a salt-bridge (permeable membrane). The current-voltage ($I-V$) characteristics of the oxidation and reduction (redox) reactions occurring at the electrodes is described by Butler-Volmer equations[14] based on standard cell potential ($\mu_{th}$), exchange current density ($J_0$) and Tafel slope ($\beta$) [13]. An EC cell can be described by a single-diode characterized by effective threshold voltage ($\mu_{th} = |\mu_{red}| + |\mu_{ox}|$), effective exchange current density ($J_0$) and effective Tafel slope ($\beta$), see supplementary material (SM) for the derivation. Assuming the resistance of the solution/electrolyte is negligible (i.e., $R_{sol} \rightarrow 0$), we obtain the following $J-V$ relationship for the EC cell:

$$J_{ec} = J_{0,ec} \exp\left(\frac{V_{cell}}{\beta}\right) = J_{0,ec} \exp\left(\frac{V_{ec}}{K\beta}\right), \quad (1)$$

where, the voltages across one cell and across a system of $K$ cells are related by $V_{cell} = V_{ec}/K$. Moreover, the effective parameters of the EC are given by,

$$J_{0,ec} \equiv J_{0,1}^{\frac{\beta_1}{\beta}} J_{0,2}^{\frac{\beta_2}{\beta}} \exp\left(\frac{-\mu_{th}}{\beta}\right) \quad (2)$$

$$\beta \equiv \beta_1 + \beta_2 \quad (3)$$

The effective exchange current density can be perceived as the weighted average of the exchange current densities of individual electrodes. An illustrative I-V characteristics for $K = 1$ system for electrolysis of water is shown in Fig. 2.

### B. I-V Characteristics of PV systems.

The general PV module is constructed from '$M$' number of series-connected cells. Each cell may have a single junction (SJ) or multi-junction (MJ) with '$N$' subcells. The module can be bifacial (with albedo, $R$) or it can be illuminated by a solar concentrator ($S$). As shown in Ref. 15, the $J-V$ relationship of a module PV of tandem cells is given by

$$J_M(V_M) = -J_{sc} + J_{0,MJ} \exp\left(\frac{qV_M}{MNk_BT}\right) \quad (4)$$

where $q$ is electronic charge, $k_B$ is Boltzmann constant, $T$ is the device temperature, $J_{sc}$ is the photocurrent which is a function of $S$ and $R$, and $J_{0,MJ}$ is the 'reverse saturation current' of a tandem cell, characterized by a set of bandgaps, $E_{g,i}$ where $i = 1$ to $N$. Substituting $N = 1$ gives us the $J-V$ relationship of a module photovoltaic comprising of series-connected single-junction solar cells. Further, substituting $N = 1$ and $M = 1$ would yield the standard $J-V$ characteristics of a single-junction solar cell, as shown in Fig. 2. Eq. (4) also describes the performance of a bifacial cell with front-side intensity, $S$, and an albedo, $R$, defined by the fraction of light reflected off ground and incident on the back surface of the cell. Eq. (4) is derived in Sec. S2 of the supplementary material.

For a given PV – EC system, we can now find an operating point $(V_{op}, J_{op})$ by solving for $I_{ec} = I_{pv} \Rightarrow A_{ec} J_{ec}(V_{op}) = A_{pv} J_M(V_{op})$, as shown in Fig. 2. Note that the ratio of cell areas ($AF = A_{pv}/A_{ec}$) is another system parameter which will appear in the discussions later. A coupling loss described by the difference in maximum power of PV and the operating power should be taken into consideration while analyzing the system.

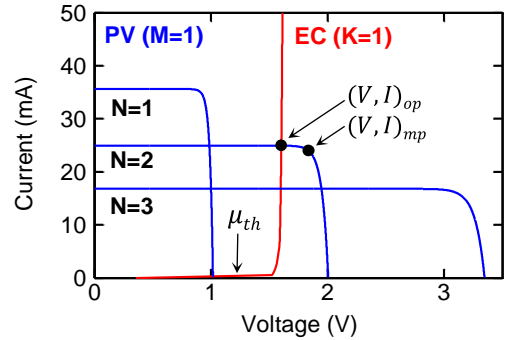

Fig. 2: I-V characteristics of PV ($M = 1, N = 1,2,3$) and EC ($K = 1$). The operating point is different from the maximum power point of PV.

Intuitively, when the EC is operated at the MP of the PV (i.e., $(V,J)_{op} = (V,J)_{mp}$), the coupling is 100%, and the system is optimized. Therefore, for the global design and optimization of the PV-EC system, we will choose $(V,J)_{op} = (V,J)_{mp}$ constraint so that $\eta_c = 1$. We can numerically find the exact solution for $(V,J)_{op}$ as explained in SI.

Since current remains almost constant for $V \leq V_{mp}$, hence, we assume that $J_{mp} \sim J_{sc}$. This allows us to use the analytical form of $J_{sc}$ as given by the following equation.

$$J_{op} = J_{sc,N} = J_{sc,top} = S J_{sun}(1 - \alpha E_{g,top}) \quad (5)$$

Further, the MP voltage of the PV can be analytically expressed as follows [3]:

$$V_{mp} = \left(\frac{MN}{q}\right)\left[E_{g,av}\left\{1 - \left(\frac{T_D}{T_S}\right)\left(\frac{E_{g,top}}{E_{g,av}}\right)\right\} - k_B T_D \left(\ln\left(\frac{\Omega_D}{S\Omega_S}\right)\right)\right] \quad (6)$$

$$E_{g,av} = \left(\frac{E_0}{N}\right) + \left(\frac{(N-1)[\alpha(1+R)E_0 - R + 1]}{2\alpha N}\right) \quad (7)$$

$$E_{g,top} = \left(\frac{N-1}{\alpha N}\right) + \left(\frac{\alpha(1+R)E_0 - R}{\alpha N}\right) \quad (8)$$

where $q$ is the electronic charge, $E_{g,av}$ (Eq. (7)) is the arithmetic mean of the bandgaps, $E_{g,top}$ (Eq. (8)) is the bandgap of the topmost subcell, $T_D$ is the device temperature, $T_S$ is the temperature of the sun, $\Omega_D$ is the emission angles of the device and $\Omega_S$ is the angle subtended by sun on the device. Equations (6-9) are constrained by $N \leq (1 + R^{-1})$ as mentioned in Ref. (15).

### C. Optimum PV-EC System.

Hence, using Eq. (1) for $K$ electrochemical cells in series, the voltage at point of operation is given by the following Eq. (9)

$$V_{op} = V_{mp} = V_{ec} = K\beta \ln\left(\frac{J_{ec}}{J_{0,ec}}\right) \quad (9)$$

Next, we substitute $V_{mp}$ from Eq. (6) and $J_{sc,N}$ from Eq. **(5)** into Eq. (9) to arrive at the key equation of this paper:

$$K\beta \ln\left[\left(\frac{SJ_{s0}(1 - \alpha E_{g,top})}{J_{0,ec}}\right) AF\right]$$

$$= \left(\frac{MN}{q}\right)\left[E_{g,av}\left\{1 - \left(\frac{T_D}{T_S}\right)\left(\frac{E_{g,top}}{E_{g,av}}\right)\right\} - k_B T_D \left(\ln\left(\frac{\Omega_D}{S\Omega_S}\right)\right)\right] \quad (10)$$

As mentioned earlier, $AF = A_{pv}/A_{ec}$. Eq. (10) determines the optimum parameters $(M, N, E_0)$ for a given EC system $(K, \mu_{th}, J_{0,ec}, \beta)$ and particular values of S and R. Note that $E_{g,top}$ and $E_{g,av}$ are functions of $E_0$, the smallest bandgap of the tandem cell. Therefore, for a set value of $(M, N, S, R)$ for the PV and given EC system, one can solve for $E_0$ from Eq. (10) for an optimal design. As we will show later, for a given EC system, a global maximum system efficiency requires: (i) co-optimization of $(M, E_0)$ at given PV module with $(N, S, R)$, or (ii) co-optimization of $(N, E_0)$ for given tandem $(M, S, R)$.

Since we find the point of operation, i.e., the intersection of $I - V$ characteristics of PV and EC for maximum power output, Eq. (10) provides the optimum parameters for system design. These parameters can be substituted in the following definition of overall system efficiency, to achieve the thermodynamic limit.

$$\eta_{sys} = \eta_{pv}\,\eta_c\eta_{ec}$$
$$= \frac{V_{mp}\,I_{mp}}{S\,P_{sun}(MA_{pv})} \times \frac{V_{op}\,I_{op}}{V_{mp}\,I_{mp}} \times \frac{K\mu_{th}\,I_{op}}{V_{op}I_{op}}$$
$$= \frac{K\,\mu_{th}\,I_{op}}{M\,S\,P_{sun}A_{pv}} = \left(\frac{K\,\mu_{th}\,J_{op}^{PV}}{M\,S\,P_{sun}}\right) \quad (11)$$

Here, $P_{sun}$ is the solar intensity reaching the PV system (~1kW/m$^2$ for AM1.5G) and $(V, I)_{mp}$ is the maximum power point of the PV module. The power required to initiate the electrochemical process at the thermodynamic equilibrium potential $\mu_{th}$ is $\mu_{th} I_{op}$. The factor $K$ accounts for the number of ECs in series. The losses in PV and EC are taken into account with their respective definitions of efficiency[13,16,17]. The coupling loss is included using the coupling efficiency, defined as the ratio of operating power over the maximum power that can be generated by the PV cell.

It is important to note that $\eta_{sys}$ also comprises of Faradaic efficiency which is assumed to be 100% in this calculation. Moreover, we use equilibrium potential (lower heating value) of the reaction and not the thermoneutral potential (higher heating value) because equilibrium potential gives an upper bound to theoretical system efficiency.

### 3. Results and discussions.

#### A. System setup and basic operation

As described in the section above, the PV system can be configured as a module consisting of series-connected single junction or multi-junction (tandem) cells. For an illustrative example, we take a water-splitting cell as the electrochemical system (load). The parameters that define a water-splitting experiment are $\mu_{th} = 1.23\,V$, $J_0 = 4.06 \times 10^{-36}$ mA/cm$^2$, $\beta = 70$ mV/decade[13].

For an intuitive understanding of the numerical optimization process, consider a single-MJ cell ($M = 1, N = 1,2,3$) in a PV system optimized for maximum efficiency (i.e., PV optimized), as shown in Fig. 2. We find that this double-junction PV provides the best coupling to the water-splitting EC ($K = 1$) and the highest system efficiency. This is because the point of operation $(V, I)_{op}$ is closest to $(V, I)_{mp}$. Fig. 2 shows that current is negligible at the point of intersection of the I-V of a single junction cell and the EC. For a triple junction ($N = 3, M = 1$) tandem cell, the overall efficiency is also lower than that of double junction ($N = 2, M = 1$) cell due to poorer coupling efficiency (current matching). This analysis implies that the optimum system efficiency depends on the number of subcells ($N$) in the tandem PV as well as the number of series-connected cells ($M$) in the module.

#### B. PV-EC system limit: comparison with literature

To illustrate the power of Eq. (10) in defining the thermodynamic limits of a variety of systems, Fig. 3 compares the experiments presented in the literature[8–11] with the thermodynamically limited efficiencies. The thermodynamic limits are calculated using Eq. (11) for the specific PV (i.e., $[E_g], S, R, M, N$) and EC setup used in the respective references. There are several highlights: the analytical results (up-triangle) are in almost perfect agreement with numerical solution (open circles), demonstrating the validity of the results. The considerable gap between thermodynamic limit and efficiency achieved in the laboratories show that there is room for considerable improvement and opportunity to quantify and reduce losses in practical systems.

One of the reasons for the gap between theoretical limits and laboratory results is that the experimental groups often use the cells that are readily available, but the bandgaps may not be optimum. We can calculate the global maximum efficiency for the same $(M, N, K, S, R)$ but with an optimum set of bandgaps, shown by green symbols using Eqs. (10),(11). The optimization proceeds as follows. For a given combination of $(M, N)$, the maximum system efficiency varies with $R$. In fact, this efficiency is attained for an optimum bandgap of the lowest bandgap subcell $(E_0)$, which determines the set of bandgaps of the multi-junction cell.

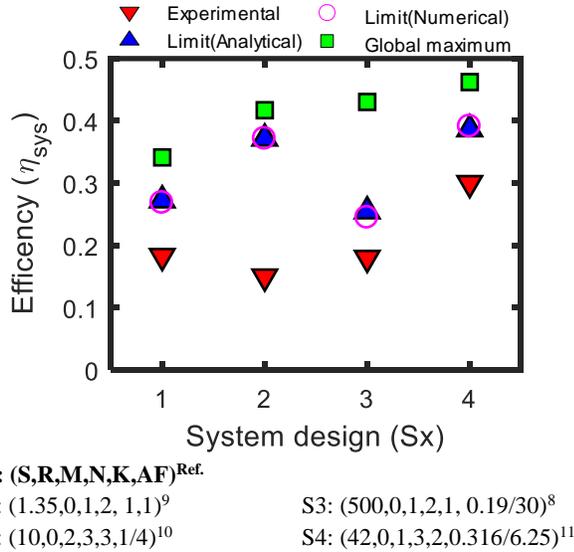

**Sx: (S,R,M,N,K,AF)[Ref.]**
S1: (1.35,0,1,2, 1,1)[9]     S3: (500,0,1,2,1, 0.19/30)[8]
S2: (10,0,2,3,3,1/4)[10]     S4: (42,0,1,3,2,0.316/6.25)[11]

Fig. 3: Comparison of reported solar to hydrogen efficiencies with the thermodynamic limit calculated analytically and numerically. Global maximum gives the best efficiency for that particular system with an optimum set of bandgaps.

### C. PV-EC system limit: $K = 1$ case

For best system efficiency, $E_0$ varies with $R$. This is evident from Fig. 4(a) and Fig. S5, which have (1,2) and (2,1) as their respective combinations of $(M, N)$. The contour plot in Fig. 4(a) distinctly shows the effect of increasing albedo $(R)$ on overall efficiency. From Fig. 4(b), we also realize that $\eta_{sys}$ for a tandem PV increases from ~33% $(R = 0)$ to ~50% $(R = 1)$ (50% increase). Similar improvements are also expected for other combinations of $M, N, R$.

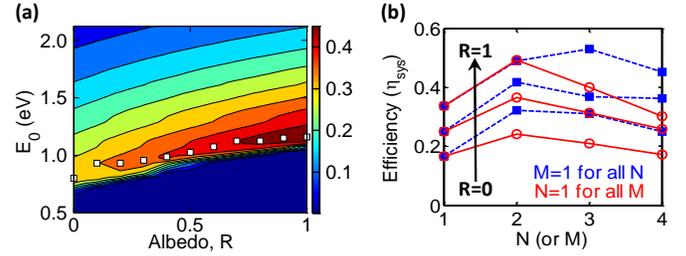

Fig. 4: (a) System efficiency as a function of $E_0$ and $R$ for $N = 2, M = 1, K = 1$ – indicating an optimum in $E_0$. (b) System efficiency increases with higher albedo and varies with N (or M). Of course, each point in this plot has a corresponding optimum $E_0$.

### D. PV-EC system limit: general case

If we revisit Eq. (10), we observe that, for $N$-junction tandem, we cannot independently set both $M$ and $K$ for an optimized design. In fact, $(M/K)$-ratio would be another optimization parameter for maximizing $\eta_{sys}$.

In most practical cases, for example on rooftops or solar farms, single junction solar cells are used. Therefore, let us first study the optimum combination of $(M, K)$ for a module of SJ $(N = 1)$ solar cells connected to a $K$- cell electrochemical system. For any SJ cell $E_g$ and known EC, one can readily calculate $(M/K)$ for optimum design using Eq. (10). The corresponding $\eta_{sys}$ is found from Eq. (11). The optimum $\eta_{sys}$ and the corresponding $(M/K)$ are shown as a function of $E_g$ in Fig. 5. For a water-splitting EC system, $\eta_{max} \sim 26.46$ % for $M/K \sim 1.67 \approx 8/5$, implying that an optimum combination of 8 SJ cells in series with 5 EC cells will yield the best overall-system efficiency. Further, this efficiency is achieved at $E_g = 1.33$ eV, which in fact is the optimum SJ PV bandgap. This is a significant new result which can be explained as follows.

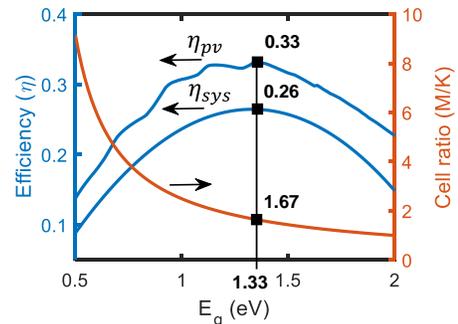

Fig 5: Variation of efficiency $(\eta)$ and cell ratio $(M/K)$ with $E_g$ (bandgap of single junction cell). Notice that maximum $\eta_{sys}$ occurs at $1.33\ eV$ which is also the optimum bandgap of best $\eta_{pv}$ achievable.

Due to logarithmic change in $V_{op}$ with current (see Eq. (9)), the EC efficiency $\eta_{ec}$ do not changes significantly as long as the change in current is relatively small (i.e., $S$ is essentially a constant). Now, with constant $\eta_{ec}$ and $\eta_c = 1$, it is obvious that the system $\eta_{sys}$ will maximize when $\eta_{pv}$ is maximum. The difference in $\eta_{pv}$ and $\eta_{sys}$ arises due to kinetic losses in EC which are incorporated in $\eta_{ec}$. Therefore, choosing an $(M/K)$-ratio so as to couple optimum-PV to the EC will indeed give the optimum system design. While we have explained the result in the context of SJ-PV and EC coupling, this analysis also holds for to tandem-PV and EC coupling, see below.

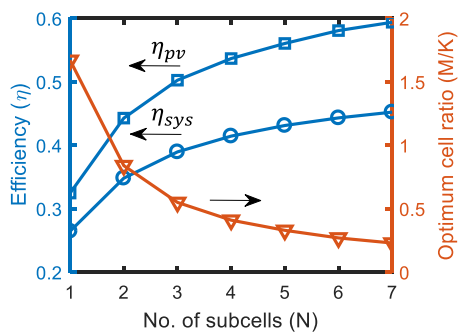

Fig. 6: Variation of efficiency ($\eta_{pv}, \eta_{sys}$) and optimum cell ratio $(M/K)$ with number of subcells ($N$) in a multi-junction cell. For every $N$, the corresponding optimum cell ratio gives the thermodynamic limit of system efficiency.

Fig. 6 shows optimum $\eta_{pv}$ and overall optimized $\eta_{sys}$ for $N$-junction tandems. The corresponding optimum $(M/K)$-ratio calculated from Eq. (10) is also shown in the same plot. The $V_{mp}$ of the optimum tandem increases with $N$ which is compensated by decreasing $(M/K)$-ratio to ensure perfect coupling between the PV module and the EC cells. The system efficiency $\eta_{sys}$ increases from 26.46% to 34.82% for $N = 1$ to 2, and starts to saturate for $N > 4$. We predict the ultimate limit of $\eta_{sys} \rightarrow 52.09\%$ as $N \rightarrow \infty$ under 1-sun with no albedo ($S = 1, R = 0$).

## 4. Summary and Conclusions.

To summarize, we have developed an analytical theory to find the thermodynamic limit of solar to fuel conversion. Given $(S, R)$ and an EC $(K, \mu_{th}, J_{0,ec}, \beta)$, system efficiency is an implicit function of number of EC cells ($K$), number of subcells ($N$), number of module cells ($M$), albedo ($R$) and bandgap ($E_0$) i.e., $\eta_{sys} = f(S, R, N, M, K, E_0, AF)$. Therefore, an optimum combination of $(M, N, E_0)$ provides the thermodynamically limited maximum efficiency which is evident from Fig. 4. The analytical formulation mentioned above provides a convenient way to find this limit and the associated parameters viz. $(M/K, N, E_0)$ and provides insights into the system with respect to recent developments in using bifaciality ($R$) and concentrated light ($S$).

The difference between the global maximum and experimental values in Fig. 3 should encourage refinement and optimization of the PV-EC design. We have considered an idealized system i.e., $R_{sol} = 0$, and it is evident that overall optimized systems, with optimum $M, N, E_0$ and $K$ can achieve much higher efficiencies compared to the laboratory results reported in the literature.[18] The analytical model developed in this work can be easily used to integrate other kinds of loads[2] to the PV system. This work can also be extended to include hourly variations in solar illumination to find daily storage capabilities and location-based optimal design.

### Acknowledgements

This work is supported by the Solar Energy Research Institute for India and the U.S. (SERIIUS) funded jointly by the U.S. Department of Energy subcontract DE AC36-08G028308 and the Government of India subcontract IUSSTF/JCERDC-SERIIUS/2012.

### References

1. A. Polman, M. Knight, E. C. Garnett, B. Ehrler and W. C. Sinke, *Science (80-. ).*, 2016, **352**, aad4424-aad4424.
2. K. K. Sakimoto, A. B. Wong and P. Yang, *Science (80-. ).*, 2016, **351**, 74–77.
3. M. R. Singh, E. L. Clark and A. T. Bell, *Proc. Natl. Acad. Sci.*, 2015, 201519212.
4. M. G. Walter, E. L. Warren, J. McKone, S. W. Boettcher, Q. Mi, E. A. Santori and N. S. Lewis, *Chem. Rev.*, 2010, **110**, 6446–6473.
5. M. Gratzel, *Nat. (London, United Kingdom)*, 2001, **414**, 338–344.
6. S. Haussener, C. Xiang, J. M. Spurgeon, S. Ardo, N. S. Lewis and A. Z. Weber, *Energy Environ. Sci.*, 2012, **5**, 9922–9935.
7. G. Kim, M. Oh and Y. Park, *Sci. Rep.*, 2016, 1–9.
8. G. Peharz, F. Dimroth and U. Wittstadt, *Int. J. Hydrogen Energy*, 2007, **32**, 3248–3252.
9. S. Licht, *J. Phys. Chem. B*, 2001, **105**, 6281–6294.
10. K. Fujii, S. Nakamura, M. Sugiyama, K. Watanabe, B. Bagheri and Y. Nakano, *Int. J. Hydrogen Energy*, 2013, **38**, 14424–14432.
11. J. Jia, L. C. Seitz, J. D. Benck, Y. Huo, Y. Chen, J. W. D. Ng, T. Bilir, J. S. Harris and T. F. Jaramillo, *Energy Procedia*, 2016, **77**, 13237.
12. J. W. Ager, M. R. Shaner, K. A. Walczak, I. D. Sharp and S. Ardo, *Energy Environ. Sci.*, 2015, **8**, 2811–2824.
13. M. T. Winkler, C. R. Cox, D. G. Nocera and T. Buonassisi, *Proc. Natl. Acad. Sci. U. S. A.*, 2013, **110**, E1076–E1082.
14. V. R. Stamenkovic, B. Fowler, B. S. Mun, G. Wang, P. N. Ross, C. a Lucas, N. M. Marković, C. Song, Y. Tang, J. Lu, J. Zhang, H. Wang, J. Shen, S. Mcdermid, J. Li, P. Kozak, P. Pietrasz, B. Orr, T. Simes, A. Staff, S. Daniel and C. Lim, *Science*, 2007, **315**, 493–7.
15. M. A. Alam and M. R. Khan, 2016, **173504**, 0–5.
16. L. C. Hirst and N. J. Ekins-Daukes, *Prog. Photovoltaics Res. Appl.*, 2011, **19**, 286–293.
17. W. Shockley and H. J. Queisser, *J. Appl. Phys.*, 1961, **32**, 510–519.
18. M. R. Singh, E. L. Clark and A. T. Bell, *Phys. Chem. Chem. Phys.*, 2015, **17**, 18924–18936.

# Supplementary Information

# Thermodynamic limit of solar to fuel conversion for generalized photovoltaic-electrochemical system

M. Tahir Patel[1], M. Ryyan Khan[1], and Muhammad A. Alam[1, a)]

[1]Electrical and Computer Engineering Department, Purdue University, West Lafayette, IN-47907, USA

## S1: Equivalent circuit for an electrochemical cell

The reactions taking place at the two electrodes of an electrochemical cell can be represented by Eqs. (S1) and (S2) using Butler-Volmer equation[1–3].

$$I_{pv} = I_{ox,C} - I_{red,C}$$
$$= I_{o,C} \left[ \exp\left(\frac{V_{pv} - V_{sol} - \mu_{red}}{\beta_{Red,C}}\right) - \exp\left(-\frac{V_{pv} - V_{sol} - \mu_{red}}{\beta_{ox,C}}\right) \right] \quad (S1)$$

$$I_{pv} = I_{red,A} - I_{ox,A}$$
$$= I_{o,A} \left[ \exp\left(\frac{V_{sol} - \mu_{ox}}{\beta_{Ox,A}}\right) - \exp\left(-\frac{V_{sol} - \mu_{ox}}{\beta_{Red,A}}\right) \right] \quad (S2)$$

These equations are depicted in the form of an electrical circuit in Fig. (S1), where a pair of "diodes" represent the redox reactions in each electrode. This electrochemical circuit is connected in series with a photovoltaic (PV) cell circuit, represented on the right by a current source and a diode. Note that in the thermodynamic limit, the shunt and series resistances need not be considered[4].

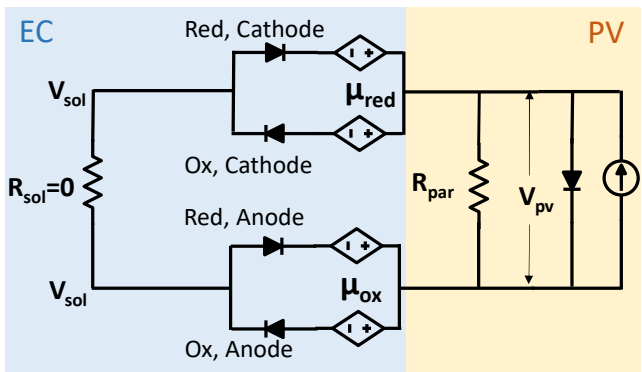

Fig. S1. Electrical equivalent circuit for EC-PV connected in series (Back-to-back-diode model).

When a reaction proceeds, one of the two back-to-back diodes at an electrode starts dominating its companion, because the first diode is forward biased, while the companion is reverse biased. In this situation, we can neglect the reverse-biased diode, because it draws negligible current as compared to the forward-biased diode. Now, we are left with two diodes, one for each electrode. The voltage across the electrochemical system is given by the sum of voltage drops across the diode, see Eq. (S3):

$$V_{ec} = \mu_{th} + \beta_1 \ln\left(\frac{J_{ec}}{J_{0,1}}\right) + \beta_2 \ln\left(\frac{J_{ec}}{J_{0,2}}\right) + J_{ec}R_{sol} \quad (S3)$$

where $R_{sol}$ is the solution resistance. Assuming an idealized case of $R_{sol} = 0$ and $\mu_{th} = \mu_1 + \mu_2$, we can rewrite Eq. (S3) in the form mentioned in the main text (Eq. (1), (2) & (3)). In the final circuit, electrochemical cell is represented as an effective single diode, as shown in Fig. (S2).

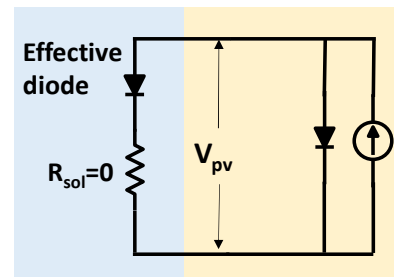

Fig. S2: Final equivalent circuit for EC-PV connected in series (Effective single diode model).

## S2: Derivation of $J - V$ relationship for PV

In the thermodynamic limit, a solar cell is described by a superposition of light and dark currents, namely

$$J(S,R,V) = -J_{ph}(S,R) + J_{dark}(V) \quad (S4)$$

where $J_{ph}$ is photocurrent generated by illumination (photons) from the sun (S) and albedo reflection (R), but is otherwise independent of voltage ($V$). The dark current ($J_{dark}$) represents the dependence of current density on voltage at zero illumination (or a dark environment). When $V = 0$, i.e., when the cell is short-circuited, $J_{dark} = 0$ and $J_{sc} = J_{ph}$.

a) Electronic mail: alam@purdue.edu

At $S$ solar concentration, the short-circuit current ($J_{sc}$) for a multi-junction solar cell is given by the following equation[4].

$$J_{sc,N} = J_{sc,top} = S J_{s0}(1 - \alpha E_{g,top}) \quad (S5)$$

where $E_{g,top}$ is the bandgap (can be found by setting $i = N - 1$ in Eq. (S9)) and $J_{sc,top}$ is the short-circuit current of the topmost subcell, respectively. The value of constants for AM 1.5G are:

$\alpha = 0.428 \, eV^{-1}$ and
$J_{s0} = \begin{cases} 83.75(1 + R/S) \text{ mA/cm}^2; \text{for } N = 1 \\ 83.75 \text{ mA/cm}^2 \quad ; \text{for } N \geq 2 \end{cases}$

The "reverse saturation current" ($J_{0,MJ}$) of the MJ cell depends on the bandgap, emission angle and the temperature of subcells. The dark current $J_{dark}$ for the MJ-cell can be written as follows[4]:

$$J_{dark} = J_{0,MJ} \exp\left(\frac{qV}{Nk_BT}\right)$$
$$= q\Omega_D \gamma \exp\left(-\frac{E_{g,av}}{k_BT}\right) \exp\left(\frac{qV}{Nk_BT}\right) \quad (S6)$$

Here, $E_{g,av}$ is the arithmetic mean of the bandgaps of subcells, see Eq. (S9) below. $\Omega_D$ is the geometric mean of emission angle of each subcell and $\gamma(E_g, T)$ is a factor that accounts for photon recycling within the subcells[4].

From Ref. (4) it can be found that for a tandem cell, the lowest bandgap ($E_0$) subcell resides at the bottom of the stack so long the constraint defined by Eq. (S7) is satisfied.

$$N \leq (1 + R^{-1}) \quad (S7)$$

Most practical systems follow this constraint. For example, even for relatively high albedo of $R = 0.5 \Rightarrow N \leq 3$. For most electrochemical reactions $\mu_{th} \leq 2$ V and $N = 3$ suffices for best system efficiencies.

A module photovoltaic system consisting of $M$ number of series-connected multi-junction solar cells has its $J - V$ characteristics described by Eq. (S8). The photocurrent density remains the same as that of a single multi-junction cell due to series connection, but the voltage across the module is divided equally across each tandem cell i.e. $V_{cell} = V_M/M$. This finally leads to the $J - V$ relation given in the main text Eq. (4) (same as Eq. (S8)).

$$J_M = -J_{sc} + J_{0,MJ} \exp\left(\frac{qV_M}{MNk_BT}\right) \quad (S8)$$

**S3: Analytical formulae related to bandgaps of tandem cell**

The set of bandgaps for a multi-junction cell, for given values of $N, R, E_0$, under the constraint shown in Eq. (S7) is given by,

$$E_i = \left(\frac{i}{\alpha N}\right) + \left(\frac{(N-i)[\alpha(1+R)E_0 - R]}{\alpha N}\right) \quad (S9)$$

where $i = 1,2,...N - 1$ is the subcell index, $\alpha = 0.428$ eV$^{-1}$ and $E_0$ is the lowest bandgap of the tandem cell. It is clear from the above expression that the set of bandgaps can be uniquely described by the parameters $N, R$ and $E_0$. This simply reflects the requirement of current-continuity for the subcells within a tandem cell. This equation is further used to find the bandgap of top subcell ($E_{g,top} = E_{N-1}$) and the average of bandgaps ($E_{g,av} = \frac{1}{N}[E_0 + \sum E_i]$), see Eqs. (7) and (8) in the main text.

**S4: V$_{mp}$ simplification**

The voltage at maximum power point ($V_{mp}$), as mentioned in the main text Eq. (6), is given by:

$$V_{mp} = \left(\frac{MN}{q}\right)\left[E_{g,av}\left\{1 - \left(\frac{T_D}{T_S}\right)\left(\frac{E_{g,top}}{E_{g,av}}\right)\right\} - k_B T_D \left(ln\left(\frac{\Omega_D}{S\,\Omega_S}\right)\right)\right] \quad (S10)$$

Since $T_D$ (300 K) >> $T_S$ (~6000 K) and $\Omega_D$ ($2\pi$ or $4\pi$), $\Omega_S$ are constants, Eq. (S10) reduces to the following simplified expression:

$$V_{mp} = \left(\frac{MN}{q}\right)(0.95\, E_{g,av} - 0.29) \quad (S11)$$

This expression of $V_{mp}$ simplifies Eq. (10) in the main text to the following:

$$K\beta \, ln\left[\left(\frac{SJ_{s0}(1 - \alpha E_{g,top})}{J_{0,ec}}\right)AF\right]$$
$$= \left(\frac{MN}{q}\right)(0.95\, E_{g,av} - 0.29) \quad (S12)$$

**S5: Comparing analytical and simulation results**

System operating point ($I_{op}, V_{op}$) can be exactly determined by numerically finding the point of intersection of the $I - V$ curves of PV and EC. Substituting the values of current in Eq. (11) of the main text can be used to calculate the efficiency.

On the other hand, the analytical Eq. (10) in the main text provides the optimum parameters of the system which can be used to find the current and hence the efficiency using Eq. (11).

Figures S3(a),(b) demonstrate that the analytical results for $M = 2,3$ (and $N = 1, K = 1$), respectively, compare well with the simulation results. The analytical results hold true for $\mu_{th} < V < V_{mp}$ and hence, $E_0$ should be large enough so that voltage across the PV can overcome the threshold voltage of EC.

a) Electronic mail: alam@purdue.edu

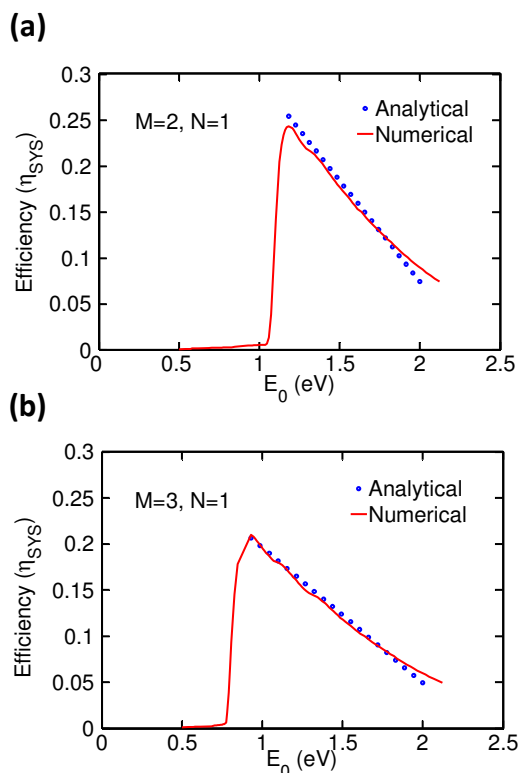

Fig. S3: Analytical vs. numerical results for (a) M=2 and (b) M=3 (K=1).

As the number of SJ solar cells increase the module voltage is simply an addition of single cell voltages, implying that volatge required from a single cell decreases, which further implies requirement of a lower value of bandgap for each cell. Fig. S4 depicts how optimum $E_0$ decreases with increase in the number of cells in the module.

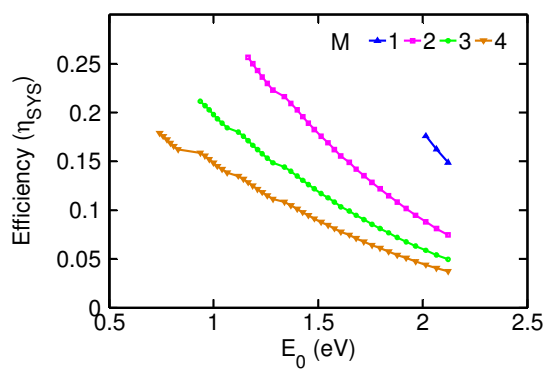

Fig. S4: For SJ module PV, the energy bandgap required for efficiently powering the EC ($K = 1$), decreases with increase in number of cells in the module.

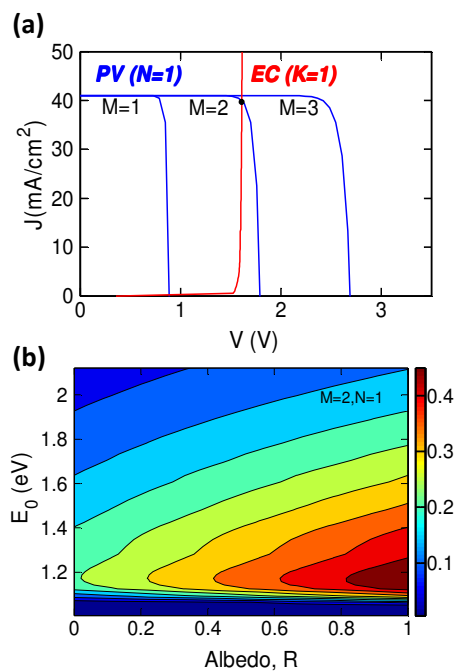

Fig. S5: (a) I-V characteristics of SJ module PV ($N = 1, M = 1,2,3$) and EC ($K = 1$) show that there is an optimum value of $M$ required for best coupling and highest system efficiency. (b) For $M = 2, N = 1, K = 1$ system efficiency is plotted for for various values of $E_0$ and $R$.

### S6: Results for a module of SJ cells

Fig. S5 shows similar results and observations as in Fig. 2 and Fig. 3 in the main text. Here, the system consists of SJ cells connected to a single cell water-splitting electrochemical system, as would be typical for many easily-implemented systems. From these figures, our two key conclusions: (a) the efficiency is an implicit function of various system parameters ($M, N, E_0, S, R$) and (b) that there is an optimum combination for best efficiency, are reinforced.

### References

1  J. A. V Butler, *Trans. Faraday Soc.*, 1924, 729–733.
2  M. Erdey-Gruz, T.; Volmer, *Z. Phys. Chem. 1930*, 1930, 203–213.
3  J. Newman, *Electrochemical Systems*, Prentice-Hall, Inc.: Englewood Cliffs, NJ, 2nd ed., 1991.
4  M. A. Alam and M. R. Khan, *Appl. Phys. Lett.*, , DOI:10.1063/1.4966137.

a) Electronic mail: alam@purdue.edu